\documentstyle[12pt]{article}
 
\begin{document} 

\baselineskip 22pt 

\begin{center}
{\Large 
\bf Decay Constant Ratios $f_{\eta_c}/f_{J/\psi}$ and
$f_{\eta_b}/f_\Upsilon$}\\
\vspace{1.0cm}
Dae Sung Hwang$^*$ and Gwang-Hee Kim$^{\dagger}$\\
{\it{Department of Physics, Sejong University, Seoul 143--747,
Korea}}\\
\vspace{2.0cm}
{\bf Abstract}\\
\end{center}

We calculate the decay constant ratios $f_{\eta_c}/f_{J/\psi}$
and $f_{\eta_b}/f_\Upsilon$. In the calculation we take into
account the mock meson structures of the mesons,
as well as the difference of the wave functions at origin of
the vector and pseudoscalar mesons studied by Ahmady and Mendel.
We find that the different spin structures of the mesons much
affect the ratios. We incorporate our results in the prediction
of the branching ratios of $B\rightarrow K\, \eta_c$.
\\

\vfill 



\noindent
$^*$e-mail: dshwang@phy.sejong.ac.kr\\
$^{\dagger}$e-mail: gkim@phy.sejong.ac.kr
\thispagestyle{empty} 
\pagebreak 

\baselineskip 22pt
The decay constants $f_{J/\psi}$ and $f_\Upsilon$ of the vector
mesons are given experimentally from the decay rates to $e^+e^-$,
but $f_{\eta_c}$ and $f_{\eta_b}$ of the pseudoscalar mesons
are lack of experimental results.
So it is necessary to calculate theoretically
the ratios $f_{\eta_c}/f_{J/\psi}$
and $f_{\eta_b}/f_\Upsilon$ for various
applications, for example, for the prediction of
$B(B^+\rightarrow K^+\, \eta_c )$ from the experimental result
$B(B^+\rightarrow K^+\, J/\psi )=
[1.02\pm 0.14]\times 10^{-3}$ \cite{cleo96a,pdg}.
It has been widely assumed that the decay constants of the vector
and pseudoscalar mesons are almost the same by considering their
wave functions at origin to be the same
and using the Van Royen-Weisskopf formula.
However, Ahmady and Mendel \cite{ahmady} calculated the ratio of
the wave functions at origin by considering the perturbation
caused by the hyperfine splitting Hamiltonian, and found that
the ratio is significantly different from unity.
Then $f_{\eta_c}$ and $f_{\eta_b}$ become very different from
$f_{J/\psi}$ and $f_\Upsilon$.
In this paper we will further take into account the influence of
the mock meson structure
\cite{royen,gi} of the mesons on the calculation of the
decay constants.
We will consider its effects on $f_{\eta_c}$, $f_{\eta_b}$,
and $f_{J/\psi}$, $f_\Upsilon$, which are originated from
the different mock meson spin structures of
the pseudoscalar and vector mesons.
We find that the effects are very important contrary to the
expectation that they are not important for the mesons composed
of two heavy quarks.
As a result, $f_{\eta_c}$ and $f_{\eta_b}$ become very close to
$f_{J/\psi}$ and $f_\Upsilon$ again,
as we will show in this paper.

The decay constants of pseudoscalar and vector mesons are defined by
\begin{equation}
<0|\, {\bar{Q}}{\gamma}^{\mu}{\gamma}_5Q'\, |M_P({\bf{K}})>=f_PK^{\mu},
\ \ \
<0|\, {\bar{Q}}{\gamma}^{\mu}Q'\, |M_V({\bf{K}},\,\varepsilon )>
=f_Vm_V{\varepsilon}^{\mu}.
\label{a4}
\end{equation}
The Van Royen-Weisskopf formula \cite{royen} for the decay constants
is given by
$f_M=\sqrt{12 / m_M}\, |\Psi_M (0)|$,
where $m_M$ and $\Psi_M (0)$ are the mass and the wave function at
origin of the meson respectively.
The Van Royen-Weisskopf formula is widely used
for the calculation of the meson decay constants.
This formula is obtained in the limit
that the spinors of the quarks inside meson are approximated
to two-component Pauli spinors \cite{royen,kaplan}.
In this paper we are interested in the $J/\psi$ and $\Upsilon$
families.
From the Van Royen-Weisskopf formula, we have
\begin{equation}
\Bigl( {f_{\eta_c}\over f_{J/\psi}}\Bigr)^2
={m_{J/\psi}\over m_{\eta_c}}
\,\, {|\Psi_{\eta_c} (0)|^2\over |\Psi_{J/\psi} (0)|^2}
=1.040\,\, {|\Psi_{\eta_c} (0)|^2\over |\Psi_{J/\psi} (0)|^2}.
\label{a7}
\end{equation}
The approximation $|\Psi_{\eta_c} (0)|=
|\Psi_{J/\psi} (0)|$ has been usually used, with which we get
$(f_{\eta_c}/f_{J/\psi})^2=1.040$.
However, recently Ahmady and Mendel calculated
$|\Psi_{\eta_c}(0)|^2/|\Psi_{J/\psi}(0)|^2$
in their interesting work \cite{ahmady} based on
the perturbation theory of quantum mechanics,
and obtained the ratio as $1.4\pm 0.1$.
By incorporating this result in (\ref{a7}),
Ahmady and Mendel obtained \cite{ahmady95}
\begin{equation}
\Bigl( {f_{\eta_c}\over f_{J/\psi}}\Bigr)^2
=1.5\pm 0.1.
\label{a8}
\end{equation}
Ahmady and Mendel also obtained
$|\Psi_{\eta_b}(0)|^2/|\Psi_{\Upsilon}(0)|^2=1.16\pm 0.06$
in their work \cite{ahmady}.
Then by using $m_{\Upsilon}=9.460$ GeV \cite{pdg} and
$m_{\eta_b}=9.445$ GeV which is calculated from
$(m_{\Upsilon}-m_{\eta_b})=(m_c/m_b)^2\, (m_{J/\psi}-m_{\eta_c})$
given by the hyperfine splitting Hamiltonian,
we obtain
\begin{equation}
\Bigl( {f_{\eta_b}\over f_{\Upsilon}}\Bigr)^2
=1.16\pm 0.06.
\label{a8b}
\end{equation}
If we had taken $|\Psi_{\eta_b} (0)|=|\Psi_{\Upsilon} (0)|$,
$(f_{\eta_b}/f_{\Upsilon})^2$ would have been given to be
1.002 from a similar equation to (\ref{a7}).
Therefore Ahmady and Mendel have taken into account the fact
that the hyperfine splitting Hamiltonian makes the wave function
at origin of the pseudoscalar meson bigger than that of the
vector meson, and then have obtained the improved results given by
(\ref{a8}) and (\ref{a8b}) which are much larger than 1.040
and 1.002 which would have been given if the wave functions at
origin had been taken to be equal.

The purpose of this paper is to improve further the results of
(\ref{a8}) and (\ref{a8b}) by taking into account
the mock meson structures of
the pseudoscalar and vector mesons, which are different by their
spin structures, as well as the difference of the wave functions
at origin which has been studied by Ahmady and Mendel.
We work
in the relativistic mock meson model of Godfrey, Isgur, and Capstick
\cite{gi,godfcg}, in which
the meson state composed of a
heavy quark $Q'$ and a heavy antiquark $\bar{Q}$
is represented by
\begin{eqnarray}
|M_P({\bf 0})>&=&{\sqrt{2m_P}}\, \int
{d^3p_{Q'}\over
(2\pi )^{3/2} {\sqrt{2E_{Q'}\, 2E_{\bar{Q}}}}}
\,\, \Phi ({\bf p}_{Q'})\,\,
{1\over {\sqrt{N_c}}}
\nonumber\\
& &\times {1\over {\sqrt{2}}}\,
[a_{\uparrow}^{\dagger}({\bf p}_{Q'},c)
b_{\downarrow}^{\dagger}({\bf p}_{\bar{Q}},{\bar{c}})
-a_{\downarrow}^{\dagger}({\bf p}_{Q'},c)
b_{\uparrow}^{\dagger}({\bf p}_{\bar{Q}},{\bar{c}})]
\, |0>
\label{f4}
\end{eqnarray}
in the meson rest frame
(where ${\bf p}_{Q'}=-{\bf p}_{\bar{Q}}$)
in which we work in this paper,
where the arrow indicates a state with spin up (down) along a
fixed axis and $c$ is the colour index which is summed.
Whereas we wrote the pseudoscalar meson state in (\ref{f4}),
we can also write
the vector meson states in the same way with the spin combinations for
the vector states, which are given by
$(\uparrow \uparrow )$,
$1/{\sqrt{2}}\, (\uparrow \downarrow +\downarrow \uparrow )$ and
$(\downarrow \downarrow )$.
In (\ref{f4}) we adopted the normalization of the creation and annihilation
operators given by
$\{ a({\bf p},s),a^{\dagger}({\bf p}',s')\} =(2\pi )^3\, 2E\, {\delta}_{ss'}
{\delta}^3({\bf p}-{\bf p}')$, and then the meson state in (\ref{f4}) is
normalized by $<M_P({\bf 0})|M_P({\bf 0})>=2m_P\, {\delta}^3({\bf 0})$,
and also in the same way for the vector meson states.
We take the momentum wave function $\Phi ({\bf{p}})$ in (\ref{f4})
as a Gaussian wave function
\begin{equation}
\Phi ({\bf{p}})={1\over ( \sqrt{\pi} \beta )^{3/2}}
e^{-{\bf{p}}^2/2{\beta}^2},\qquad
\Psi ({\bf r})=({{\beta}\over {\sqrt{\pi}}})^{3/2}
e^{-{\beta}^2{\bf r}^2/2},
\label{f2}
\end{equation}
where $\Psi ({\bf r})$ is the conjugate wave function in coordinate
space.
In Fig. 1, we display six different
inter-quark potentials of the potential models in
Refs. \cite{eich}--\cite{richard},
which were obtained by fitting the data
of the $J/\psi$ and $\Upsilon$ families (mainly their spectra).
The mean square radii $\langle r^2 {\rangle}^{1/2}$
of the $J/\psi$ and $\Upsilon$ mesons are about
$2.2\ {\rm GeV}^{-1}$ and
$1.3\ {\rm GeV}^{-1}$ \cite{apctp}, respectively,
which are in the confining
long-distant linear potential range.
Therefore, using the Gaussian wave function in (\ref{f2}) is appropriate.
In particular, since we will calculate the ratios of the decay constants
by using (\ref{f10}), the Gaussian wave function is reliable to use
in our following calculations.

Since we are concerned with the matrix elements in the left hand sides of
(\ref{a4}) with the meson states in (\ref{f4}), it is convenient to
represent the meson states by
the matrix-valued representations given by
\begin{equation}
{\Psi}_{P\,\alpha\beta}\equiv
- <0|\, Q'_{\alpha}{\bar{Q}}_{\beta}\, |M_P({\bf 0})>,\ \ \
{\Psi}_{V\,\alpha\beta}\equiv
- <0|\, Q'_{\alpha}{\bar{Q}}_{\beta}\, |M_V({\bf 0})>,
\label{f5}
\end{equation}
where $\alpha$, $\beta$ are spinor indeces.
With (\ref{f5}), the formulas in (\ref{a4}) are written as
\begin{equation}
Tr(\, {\gamma}^{0}{\gamma}_5\, {\Psi}_P\, )=f_Pm_P,\ \ \
Tr(\, {\gamma}^{\mu}{\Psi}_V\, )=f_Vm_V{\varepsilon}^{\mu}.
\label{f6}
\end{equation}
If both two quarks inside the meson are static,
the spinor combinations of $u({\bf 0}){\bar{v}}({\bf 0})$
for the pseudoscalar and vector meson states
are given respectively by \cite{falk,neubert1}
\begin{equation}
P({\bf 0},{\bf 0})=
-{1\over {\sqrt{2}}}{1+{\gamma}^0\over 2}{\gamma}^5,\ \ \
V({\bf 0},{\bf 0},\varepsilon )=
{1\over {\sqrt{2}}}{1+{\gamma}^0\over 2}\not{\varepsilon},
\label{f7}
\end{equation}
where the polarization vectors of the vector meson are given by
${\varepsilon}^{\mu}_{\pm}=(1/{\sqrt{2}})(0,1,\pm i,0)$ and
${\varepsilon}^{\mu}_3=(0,0,0,1)$.
By Lorentz boosting the static spinors we obtain
${\Psi}_P$ and ${\Psi}_V$ in (\ref{f5}) as
\begin{equation}
{\Psi}_I=
{\sqrt{2m_I}}\int {d^3p_{Q'}\over (2\pi )^{3/2}}\,
\Phi ({\bf{p}}_{Q'})
{\sqrt{N_c}\over \sqrt{2E_{Q'}\, 2E_{\bar{Q}}}}
\, {{\not{p}}_{Q'}+m_{Q'}\over {\sqrt{2m_{Q'}(m_{Q'}+E_{Q'})}}}
S_I
{-{\not{p}}_{\bar{Q}}+m_{\bar{Q}}\over
{\sqrt{2m_{\bar{Q}}(m_{\bar{Q}}+E_{\bar{Q}})}}},
\label{f9}
\end{equation}
where $I=P$ or $V$, and $S_P$ and $S_V$ are respectively
$P({\bf 0},{\bf 0})$ and
$V({\bf 0},{\bf 0},\varepsilon )$ in (\ref{f7}).
By incorporating (\ref{f9}) in (\ref{f6}),
we obtain the following formula for the
decay constants of pseudoscalar and vector mesons in the
relativistic mock meson model:
\begin{equation}
f_I=
{2{\sqrt{3}}\over {\sqrt{m_I}}}\int {d^3p\over (2\pi )^{3/2}}\,
\Phi ({\bf{p}})\Big(
{E_{Q'}+m_{Q'}\over 2E_{Q'}}\,
{E_{\bar{Q}}+m_{\bar{Q}}\over 2E_{\bar{Q}}}
{\Big)}^{1/2}
\,\Big( \, 1\, +\, a_I\,
{{\bf{p}}^2\over (E_{Q'}+m_{Q'})(E_{\bar{Q}}+m_{\bar{Q}})}\, \Big) ,
\label{f10}
\end{equation}
where $a_P=-1$ and $a_V=+1/3$.
We note that the above formulas for the decay constants
have been derived by taking the four-component spinor into
consideration, and
it is reduced to
the Van Royen-Weisskopf formula
in the two-component spinor limit
which corresponds to taking the ${\bf p}\rightarrow {\bf 0}$
limit in the last two factors of (\ref{f10}).

When the meson and quark masses are given,
$f_{\eta_c}$ and $f_{J/\psi}$ can be calculated from
(\ref{f10}) for a given value of
the parameter $\beta$ in (\ref{f2}).
We obtained them numerically
as functions of
$\beta$ by using $m_{J/\psi}=3.097$ GeV,
$m_{\eta_c}=2.979$ GeV \cite{pdg},
and $m_c=1.78$ GeV \cite{lich}.
We present the results in Fig. 2.
We performed the same calculations for
$f_{\eta_b}$ and $f_{\Upsilon}$ with
$m_{\Upsilon}=9.460$ GeV \cite{pdg}, $m_{\eta_b}=9.445$ GeV,
and $m_b=5.17$ GeV \cite{lich},
and present the results in Fig. 3.

Using the experimental values
$\Gamma (J/\psi\to e^+e^-) = 5.27\pm 0.37$ keV,
$\Gamma (\Upsilon\to e^+e^-) = 1.32\pm 0.03$ keV \cite{pdg},
and the formula \cite{neubert11}
\begin{equation}
\Gamma (V\to e^+e^-) = {4\pi\over 3}{\alpha^2\over m_V}
f_V^2c_V ,\ \ \
{\rm where}\ c_{J/\psi}={4\over 9}\ 
{\rm and}\ c_{\Upsilon}={1\over 9},
\label{a11}
\end{equation}
we get
\begin{equation}
f_{J/\psi}=406\pm 14\ {\rm MeV},\qquad
f_{\Upsilon}=710\pm 8\ {\rm MeV}.
\label{a12}
\end{equation}
Then, from (\ref{f10}) we obtain
\begin{equation}
\beta_{J/\psi} = 0.644\pm 0.022 \ {\rm GeV},\qquad
\beta_{\Upsilon} = 1.360\pm 0.016 \ {\rm GeV}.
\label{a13}
\end{equation}
In obtaining the above results, we took the quark masses as
$m_c=1.30-1.85$ GeV and $m_b=4.70-5.20$ GeV,
which cover the quark masses in the six potential models
in Refs. \cite{eich}--\cite{richard}.
We note that the errors in (\ref{a13}) came from the quark
mass ranges considered, as well as from the experimental
errors of the vector meson decay constants in (\ref{a12}).
From the results of Ahmady and Mendel \cite{ahmady},
$|\Psi_{\eta_c}(0)|^2/|\Psi_{J/\psi}(0)|^2\ =\ 1.4\pm 0.1$
and
$|\Psi_{\eta_b}(0)|^2/|\Psi_{\Upsilon}(0)|^2\ =\ 1.16\pm 0.06$,
we get
\begin{eqnarray}
\beta_{\eta_c} &=& \beta_{J/\psi} \times (1.4\pm 0.1)^{1/3}=
0.721\pm 0.042 \ {\rm GeV},
\label{a14}\\
\beta_{\eta_b} &=& \beta_{\Upsilon} \times (1.16\pm 0.06)^{1/3}
=1.428\pm 0.041 \ {\rm GeV},
\nonumber
\end{eqnarray}
since $|\Psi_M(0)|^2=(\beta_M/{\sqrt{\pi}})^3$ from (\ref{f2}).
Then by using these values of $\beta_{\eta_c}$
and $\beta_{\eta_b}$
in (\ref{f10}), we obtain
\begin{equation}
f_{\eta_c}=420\pm 52\ {\rm MeV},\qquad
f_{\eta_b}=705\pm 27\ {\rm MeV},
\label{a15}
\end{equation}
and then the ratios are given by
\begin{equation}
\Bigl( {f_{\eta_c}\over f_{J/\psi}}\Bigr)^2
=1.06\pm 0.14,\qquad
\Bigl( {f_{\eta_b}\over f_{\Upsilon}}\Bigr)^2
=0.99\pm 0.04,
\label{a16}
\end{equation}
instead of (\ref{a8}) and (\ref{a8b}).
We obtained the results (\ref{a16}) by taking
into account the different mock meson spin structures of
the pseudoscalar and vector mesons, as well as their different
values of the wave functions at origin caused by the hyperfine
splitting Hamiltonian.
We note that the difference between the results in (\ref{a16})
and those in (\ref{a8}) and (\ref{a8b}) has come from the
relativistic correction (obtained by taking the four-component
spinor into consideration) which is calculated by assuming
a Gaussian wave function for the quarkonium.
The obtained ratios in (\ref{a16})
are much close to unity compared to the
results of (\ref{a8}) and (\ref{a8b}) which were obtained by
Ahmady and Mendel \cite{ahmady,ahmady95} by taking into account
only the difference of the wave functions at origin.

The errors in the above results
(\ref{a13})--(\ref{a16}) come from the ranges of the values of
the quark masses $m_c$ and $m_b$ which we took as
$m_c=1.30-1.85$ GeV and $m_b=4.70-5.20$ GeV,
as well as the experimental errors in the values of
$f_{J/\psi}$ and $f_{\Upsilon}$ in (\ref{a12}).
In order to see the magnitude of the error in the results
induced by the sensitivity to the quark masses,
we present the results which are obtained when we use fixed
values of $m_c$ and $m_b$ in the calculation:
When we use $m_c=1.78$ GeV and $m_b=5.17$ GeV \cite{lich},
(\ref{a13}) becomes
$\beta_{J/\psi} = 0.638\pm 0.015 \ {\rm GeV},\
\beta_{\Upsilon} = 1.357\pm 0.013 \ {\rm GeV}$,
(\ref{a14})
$\beta_{\eta_c} = 0.714\pm 0.034 \ {\rm GeV},\
\beta_{\eta_b} = 1.426\pm 0.038 \ {\rm GeV}$,
(\ref{a15})
$f_{\eta_c}=424\pm 25\ {\rm MeV},\
f_{\eta_b}=709\pm 20\ {\rm MeV}$,
and finally (\ref{a16}) becomes
$(f_{\eta_c} / f_{J/\psi})^2=1.09\pm 0.07,\
(f_{\eta_b} / f_{\Upsilon})^2=1.00\pm 0.04$.
By comparing these values
with those in (\ref{a15}) and (\ref{a16}),
we can see the sensitivity of the results to the values of the
quark masses $m_c$ and $m_b$.

The decays of $B\rightarrow K\, J/\psi$ are important for the check
of the factorization hypothesis and the search of the $CP$
violation phenomena in the $B$ meson decays,
therefore there have been continuous and intensive experimental
improvements on their measurements.
Recently CLEO reported the new results \cite{cleo96a}:
\begin{eqnarray}
B(B^0\rightarrow K^0\, J/\psi )&=&
[1.15\pm 0.23({\rm stat})\pm 0.17({\rm syst})]\times 10^{-3},
\nonumber\\
B(B^0\rightarrow K^*(892)^0\, J/\psi )&=&
[1.36\pm 0.27({\rm stat})\pm 0.22({\rm syst})]\times 10^{-3},
\label{a1}\\
B(B^+\rightarrow K^*(892)^+\, J/\psi )&=&
[1.58\pm 0.47({\rm stat})\pm 0.27({\rm syst})]\times 10^{-3},
\nonumber
\end{eqnarray}
by incorporating the world average \cite{pdg}
\begin{equation}
B(B^+\rightarrow K^+\, J/\psi )=
[1.02\pm 0.14]\times 10^{-3}.
\label{a2}
\end{equation}
In connection with $B\rightarrow K\, J/\psi$,
the decays of $B\rightarrow K\, \eta_c$ have been
intensively studied theoretically
\cite{ahmady95,gourdin95,colangelo95},
since the decays are expected to be measured in near future
and by comparing the two decay modes we can get a lot of
valuable informations on the hadronic structures.
Studying the form factors phenomenologically in detail,
Gourdin $et$ $al.$ obtained the following results \cite{gourdin95}:
\begin{eqnarray}
T={\Gamma (B\rightarrow K\, \eta_c)\over
\Gamma (B\rightarrow K\, J/\psi )}
=\bar{T}\times \Bigl( {f_{\eta_c}\over f_{J/\psi}}\Bigr)^2&,&
0.957\leq \bar{T} \leq 1.259,
\nonumber\\
T^*={\Gamma (B\rightarrow K^*\, \eta_c)\over
\Gamma (B\rightarrow K^*\, J/\psi )}
=\bar{T^*}\times \Bigl( {f_{\eta_c}\over f_{J/\psi}}\Bigr)^2&,&
0.456\leq \bar{T^*} \leq 0.872.
\label{a3}
\end{eqnarray}
By incorporating our obtained decay constant ratios (\ref{a16})
in (\ref{a3}),
from (\ref{a1}) and (\ref{a2}) we predict
\begin{eqnarray}
& &B(B^+\rightarrow K^+\, \eta_c)=
\bar{T}\times [(1.11\pm 0.17)\times 10^{-3}]=
[0.90\sim 1.61]\times 10^{-3},
\nonumber\\
& &B(B^0\rightarrow K^0\, \eta_c)=
\bar{T}\times [(1.25\pm 0.32)\times 10^{-3}]=
[0.89\sim 1.98]\times 10^{-3},
\label{a1b}\\
& &B(B^+\rightarrow K^*(892)^+\, \eta_c)=
\bar{T^*}\times [(1.72\pm 0.60)\times 10^{-3}]=
[0.51\sim 2.03]\times 10^{-3},
\nonumber\\
& &B(B^0\rightarrow K^*(892)^0\, \eta_c)=
\bar{T^*}\times [(1.48\pm 0.39)\times 10^{-3}]=
[0.50\sim 1.63]\times 10^{-3}.
\nonumber
\end{eqnarray}
In obtaining (\ref{a1b}) we combined the errors in (\ref{a16}),
(\ref{a1}) and (\ref{a2}) by root mean square.
For the predictions in (\ref{a1b}) we used the range
in (\ref{a3}) of the values
of $\bar{T}$ and $\bar{T^*}$ obtained by Gourdin $et$ $al.$
\cite{gourdin95}.
However, $\bar{T}$ and $\bar{T^*}$ are very dependent
(especially $\bar{T^*}$) on the model for the form factors of
$(B\to K)$ and $(B\to K^*)$.
Ahmady and Mendel applied the heavy quark effective theory
and obtained $\bar{T}=1.12$ and $\bar{T^*}=0.27$ \cite{ahmady95},
whose $\bar{T}$ is inside, but $\bar{T^*}$ is outside of the range
in (\ref{a3}).
Particularly, Deshpande and Trampetic \cite{deshpande} emphasized
that the value of $\bar{T^*}$ is very much dependent on the model
adopted and pointed out that the measurements of the decay
$B\rightarrow K^*\, \eta_c$ will provide a valuable
criterion for the model for the hadronic form factors.

In conclusion,
we calculated the decay constant ratios $f_{\eta_c}/f_{J/\psi}$
and $f_{\eta_b}/f_\Upsilon$ by taking into
account the mock meson structures of the mesons,
as well as the difference of the wave functions at origin of
the pseudoscalar and vector mesons.
The results have been significantly affected by
the different mock meson spin structures of the mesons.
We incorporated the obtained ratios of the decay constants
in the prediction of the branching ratios of
the $B\rightarrow K\, \eta_c$ decays.
\\

\vspace*{1.0cm}

\noindent
{\em Acknowledgements} \\
\indent
This work was supported
in part by the Basic Science Research Institute Program,
Ministry of Education, Project No. BSRI-95-2414,
and in part by Non-Directed-Research-Fund,
Korea Research Foundation 1996.\\

\pagebreak

\pagebreak

\noindent
{\large\bf
Figure Captions}\\

\noindent
Fig. 1. The inter-quark potentials of the potential models in
\cite{eich}--\cite{richard}.
The radial distance of the horizontal axis is in the unit of
${\rm{GeV}}^{-1}$ (1 ${\rm{GeV}}^{-1}$ = 0.197 fm),
and the potential energy of the vertical axis is in the unit
of GeV.\\

\noindent
Fig. 2.
$f_{\eta_c}$ and $f_{J/\psi}$
as functions of the parameter $\beta$.\\

\noindent
Fig. 3.
$f_{\eta_b}$ and $f_{\Upsilon}$
as functions of the parameter $\beta$.\\

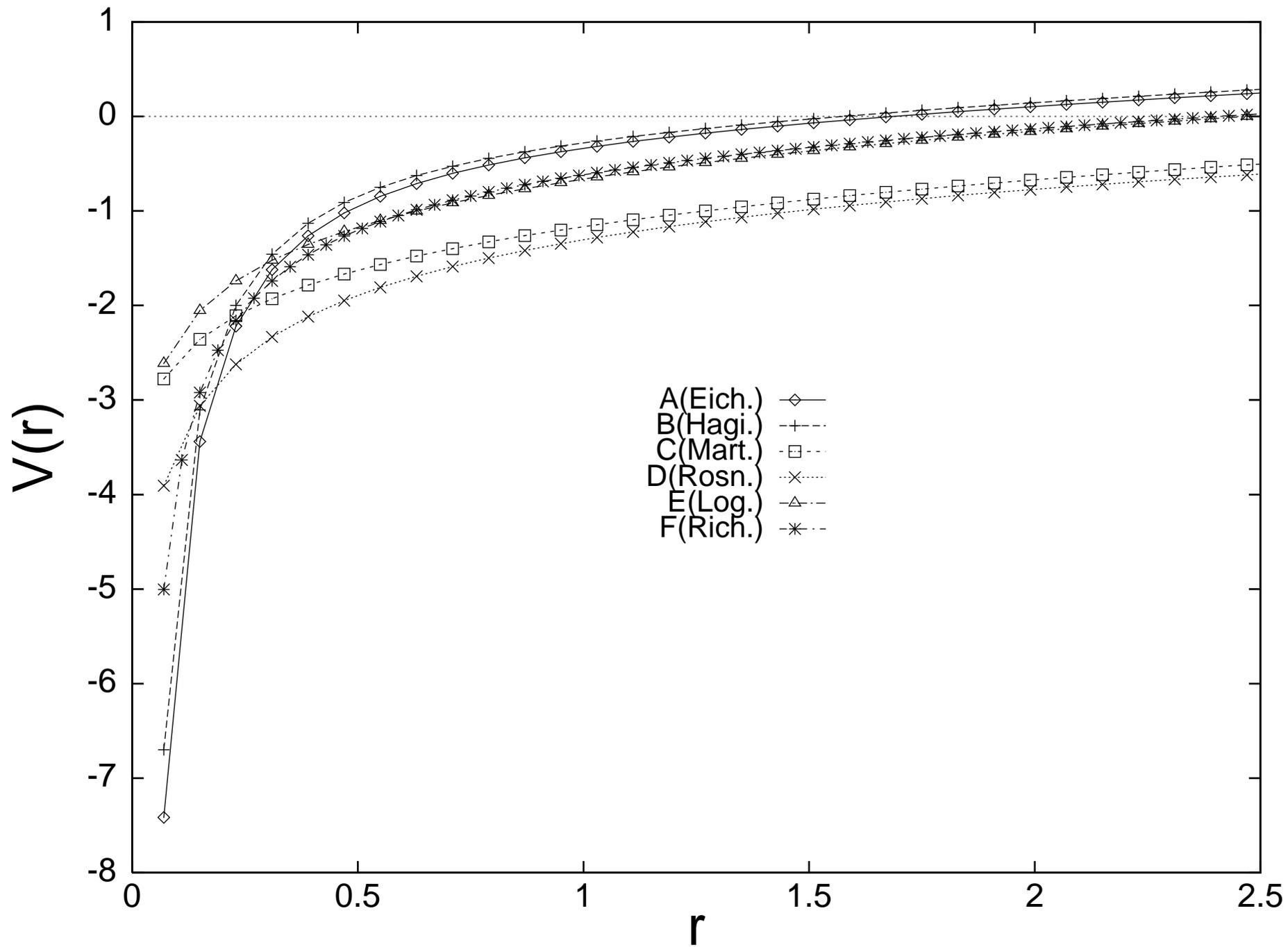
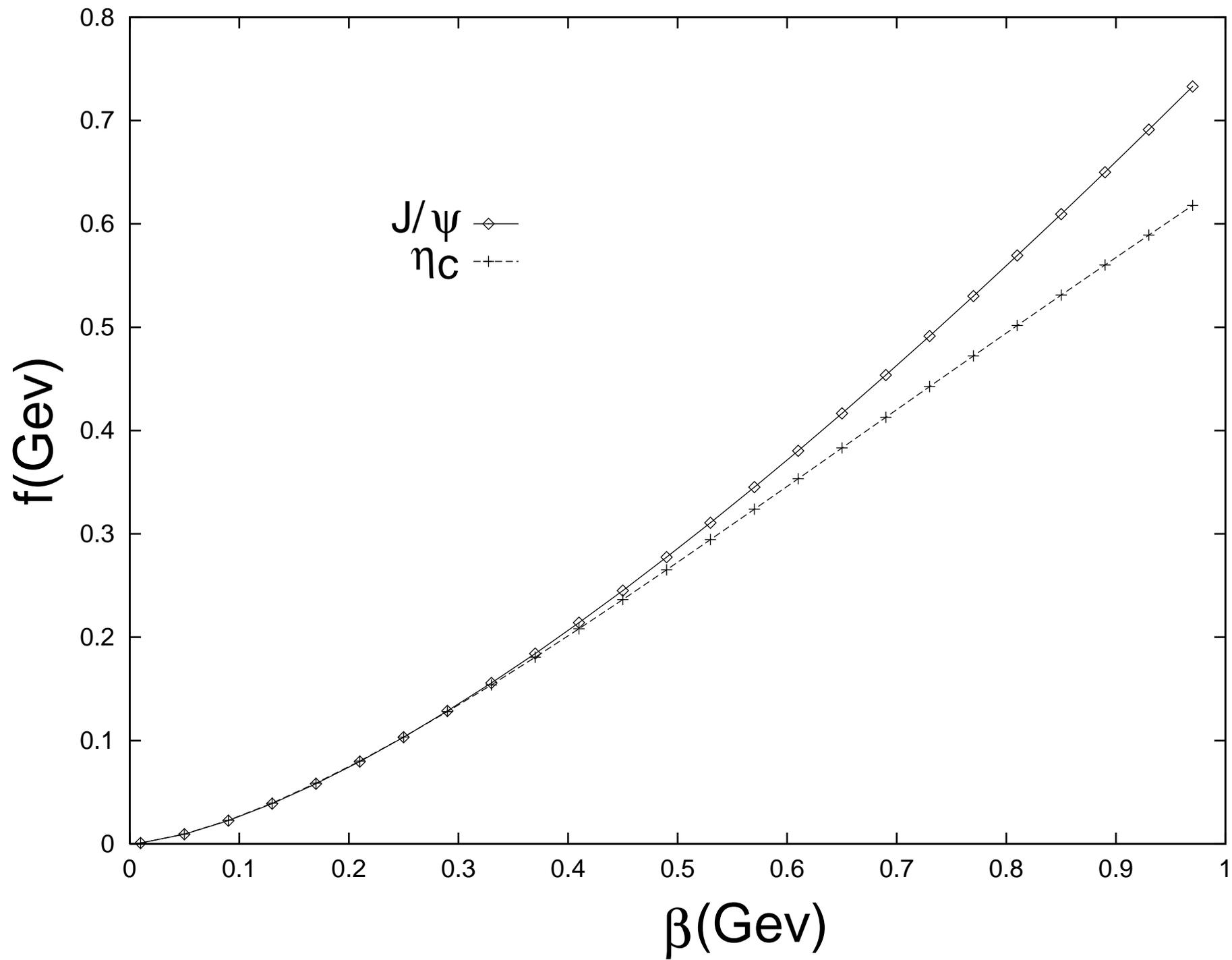
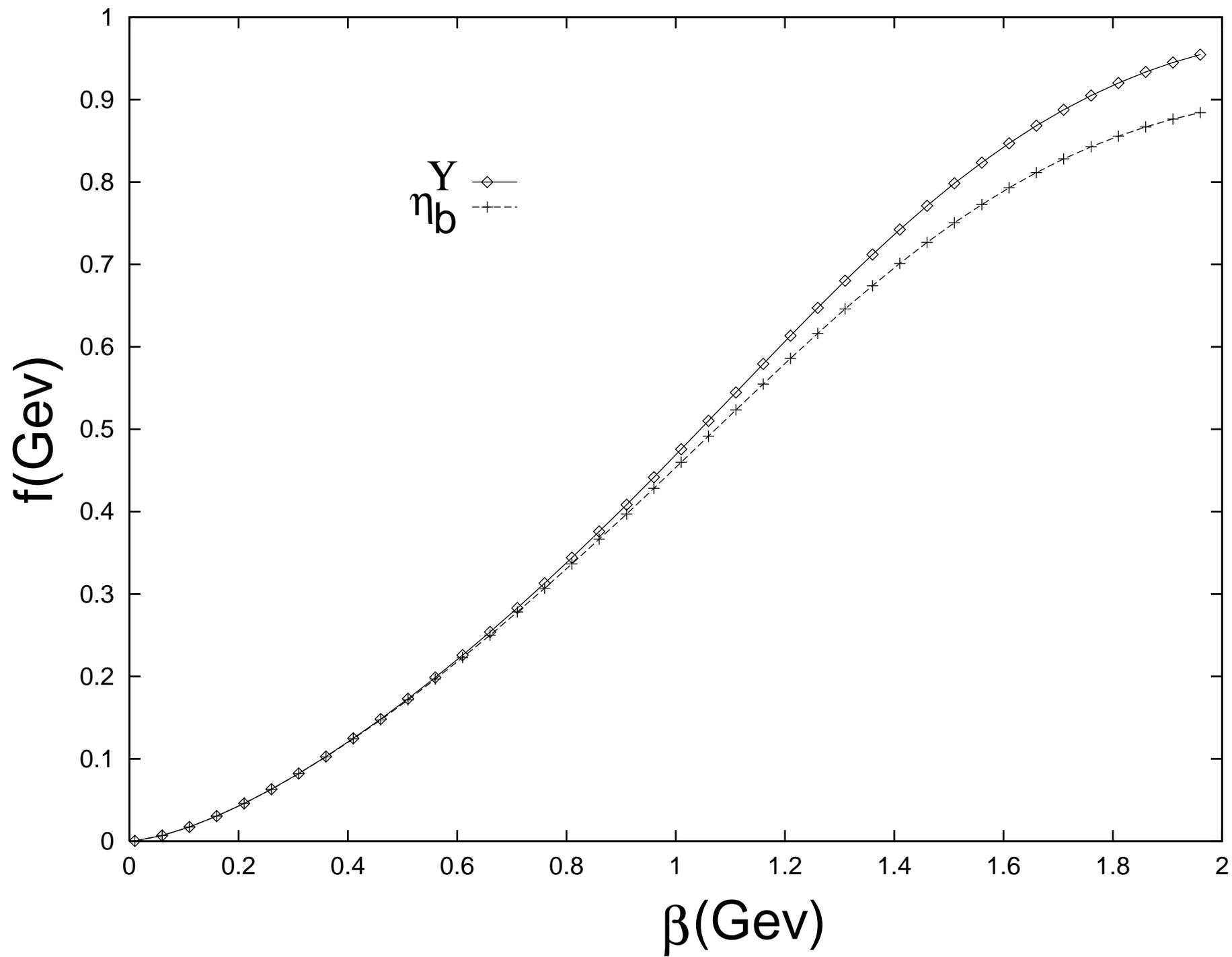
\end{document}